\documentclass[aps,prl,twocolumn,showpacs,superscriptaddress,preprintnumbers]{revtex4}
\usepackage{graphicx}
\usepackage{subfigure}
\usepackage{amsmath}
\def\vermark{} % v0.71->v0.80: for submission

\def\bsjf0BF{\ensuremath{(1.16^{+0.31}_{-0.19}(\mathrm{stat.})^{+0.15}_{-0.17}(\mathrm{syst.})
^{+0.26}_{-0.18}(N_{B_s^{(*)}\bar B_s^{(*)}})) \times 10^{-4}}}
\def\bsjfxBF{\ensuremath{(0.34^{+0.11}_{-0.14}(\mathrm{stat.})^{+0.03}_{-0.02}(\mathrm{syst.})
^{+0.08}_{-0.05}(N_{B_s^{(*)}\bar B_s^{(*)}})) \times 10^{-4}}}

\begin{document}
\makeatletter
\newcommand{\ps@mytitle}{%
  \renewcommand{\@oddhead}{{\vermark}\hfil}%
}
\makeatother

\title{Observation of $B_s^0\to J/\psi f_0(980)$ and 
Evidence for $B_s^0\to J/\psi f_0(1370)$}
\affiliation{Budker Institute of Nuclear Physics, Novosibirsk}
\affiliation{Faculty of Mathematics and Physics, Charles University, Prague}
%%%\affiliation{Chiba University, Chiba}
\affiliation{University of Cincinnati, Cincinnati, Ohio 45221}
%%%\affiliation{Department of Physics, Fu Jen Catholic University, Taipei}
\affiliation{Justus-Liebig-Universit\"at Gie\ss{}en, Gie\ss{}en}
\affiliation{Gifu University, Gifu}
%%%\affiliation{The Graduate University for Advanced Studies, Hayama}
\affiliation{Gyeongsang National University, Chinju}
\affiliation{Hanyang University, Seoul}
\affiliation{University of Hawaii, Honolulu, Hawaii 96822}
\affiliation{High Energy Accelerator Research Organization (KEK), Tsukuba}
%%%\affiliation{Hiroshima Institute of Technology, Hiroshima}
\affiliation{University of Illinois at Urbana-Champaign, Urbana, Illinois 61801}
\affiliation{Indian Institute of Technology Guwahati, Guwahati}
%%%\affiliation{Institute of High Energy Physics, Chinese Academy of Sciences, Beijing}
\affiliation{Institute of High Energy Physics, Vienna}
\affiliation{Institute of High Energy Physics, Protvino}
%%%\affiliation{Institute of Mathematical Sciences, Chennai}
%%%\affiliation{INFN - Sezione di Torino, Torino}
\affiliation{Institute for Theoretical and Experimental Physics, Moscow}
\affiliation{J. Stefan Institute, Ljubljana}
\affiliation{Kanagawa University, Yokohama}
\affiliation{Institut f\"ur Experimentelle Kernphysik, Karlsruher Institut f\"ur Technologie, Karlsruhe}
\affiliation{Korea Institute of Science and Technology Information, Daejeon}
\affiliation{Korea University, Seoul}
%%%\affiliation{Kyoto University, Kyoto}
\affiliation{Kyungpook National University, Taegu}
\affiliation{\'Ecole Polytechnique F\'ed\'erale de Lausanne (EPFL), Lausanne}
\affiliation{Faculty of Mathematics and Physics, University of Ljubljana, Ljubljana}
\affiliation{University of Maribor, Maribor}
\affiliation{Max-Planck-Institut f\"ur Physik, M\"unchen}
\affiliation{University of Melbourne, School of Physics, Victoria 3010}
\affiliation{Nagoya University, Nagoya}
%%%\affiliation{Nara University of Education, Nara}
\affiliation{Nara Women's University, Nara}
\affiliation{National Central University, Chung-li}
\affiliation{National United University, Miao Li}
\affiliation{Department of Physics, National Taiwan University, Taipei}
\affiliation{H. Niewodniczanski Institute of Nuclear Physics, Krakow}
\affiliation{Nippon Dental University, Niigata}
\affiliation{Niigata University, Niigata}
\affiliation{University of Nova Gorica, Nova Gorica}
\affiliation{Novosibirsk State University, Novosibirsk}
\affiliation{Osaka City University, Osaka}
%%%\affiliation{Osaka University, Osaka}
\affiliation{Panjab University, Chandigarh}
%%%\affiliation{Peking University, Beijing}
%%%\affiliation{Princeton University, Princeton, New Jersey 08544}
\affiliation{Research Center for Nuclear Physics, Osaka}
\affiliation{RIKEN BNL Research Center, Upton, New York 11973}
\affiliation{Saga University, Saga}
\affiliation{University of Science and Technology of China, Hefei}
\affiliation{Seoul National University, Seoul}
%%%\affiliation{Shinshu University, Nagano}
\affiliation{Sungkyunkwan University, Suwon}
\affiliation{School of Physics, University of Sydney, NSW 2006}
\affiliation{Tata Institute of Fundamental Research, Mumbai}
\affiliation{Excellence Cluster Universe, Technische Universit\"at M\"unchen, Garching}
%%%\affiliation{Toho University, Funabashi}
\affiliation{Tohoku Gakuin University, Tagajo}
\affiliation{Tohoku University, Sendai}
\affiliation{Department of Physics, University of Tokyo, Tokyo}
\affiliation{Tokyo Institute of Technology, Tokyo}
\affiliation{Tokyo Metropolitan University, Tokyo}
\affiliation{Tokyo University of Agriculture and Technology, Tokyo}
%%%\affiliation{Toyama National College of Maritime Technology, Toyama}
\affiliation{CNP, Virginia Polytechnic Institute and State University, Blacksburg, Virginia 24061}
\affiliation{Wayne State University, Detroit, Michigan 48202}
%%%\affiliation{Yamagata University, Yamagata}
\affiliation{Yonsei University, Seoul}
% \author{I.~Adachi}\affiliation{High Energy Accelerator Research Organization (KEK), Tsukuba} % KEK
% \author{K.~Adamczyk}\affiliation{H. Niewodniczanski Institute of Nuclear Physics, Krakow} % Krakow
  \author{J.~Li}\affiliation{Seoul National University, Seoul} % Seoul
  \author{H.~Aihara}\affiliation{Department of Physics, University of Tokyo, Tokyo} % Tokyo
  \author{K.~Arinstein}\affiliation{Budker Institute of Nuclear Physics, Novosibirsk}\affiliation{Novosibirsk State University, Novosibirsk} % BINP
% \author{Y.~Arita}\affiliation{Nagoya University, Nagoya} % Nagoya
% \author{T.~Aso}\affiliation{Toyama National College of Maritime Technology, Toyama} % Toyama
  \author{V.~Aulchenko}\affiliation{Budker Institute of Nuclear Physics, Novosibirsk}\affiliation{Novosibirsk State University, Novosibirsk} % BINP
  \author{T.~Aushev}\affiliation{\'Ecole Polytechnique F\'ed\'erale de Lausanne (EPFL), Lausanne}\affiliation{Institute for Theoretical and Experimental Physics, Moscow} % ITEP
  \author{T.~Aziz}\affiliation{Tata Institute of Fundamental Research, Mumbai} % Tata
  \author{A.~M.~Bakich}\affiliation{School of Physics, University of Sydney, NSW 2006} % Sydney
  \author{V.~Balagura}\affiliation{Institute for Theoretical and Experimental Physics, Moscow} % ITEP
% \author{Y.~Ban}\affiliation{Peking University, Beijing} % Peking
  \author{E.~Barberio}\affiliation{University of Melbourne, School of Physics, Victoria 3010} % Melbourne
% \author{A.~Bay}\affiliation{\'Ecole Polytechnique F\'ed\'erale de Lausanne (EPFL), Lausanne} % Lausanne
% \author{I.~Bedny}\affiliation{Budker Institute of Nuclear Physics, Novosibirsk}\affiliation{Novosibirsk State University, Novosibirsk} % BINP
  \author{K.~Belous}\affiliation{Institute of High Energy Physics, Protvino} % Protvino
% ...
% Karlsruhe
% \author{N.~Zwahlen}\affiliation{\'Ecole Polytechnique F\'ed\'erale de Lausanne (EPFL), Lausanne} % Lausanne
% \author{O.~Zyukova}\affiliation{Budker Institute of Nuclear Physics, Novosibirsk}\affiliation{Novosibirsk State University, Novosibirsk} % BINP
\collaboration{The Belle Collaboration}

\begin{abstract}
We report the first observation of $B_s^0\to J/\psi f_0(980)$ and
first evidence for $B_s^0\to J/\psi f_0(1370)$, which are CP
eigenstate decay modes.  These results are obtained from
$121.4\;\mathrm{fb}^{-1}$ of data collected at the $\Upsilon(5S)$
resonance with the Belle detector at the KEKB $e^+e^-$ collider.  We
measure the branching fractions $\mathcal{B}(B_s^0\to J/\psi
f_0(980);f_0(980)\to\pi^+\pi^-)=\bsjf0BF $ with a significance of $8.4
\sigma$, and $\mathcal{B}(B_s^0\to J/\psi
f_0(1370);f_0(1370)\to\pi^+\pi^-)=\bsjfxBF $ with a significance of
$4.2 \sigma$.  The last error listed is due to uncertainty in the
number of produced $B_s^{(*)}\bar B_s^{(*)}$ pairs.
\end{abstract}
 % Decays of bottom mesons

\maketitle
\thispagestyle{mytitle}
\markright{\vermark}

The $b\to c \bar c s$ process $B_s^0\to J/\psi \phi$, which has a
relatively large branching fraction, has been used to extract the
$B_s^0$ decay width difference $\Delta\Gamma$ and $CP$-violating phase
$\beta_s$ from time-dependent analyses~\cite{Aaltonen:2007he}.  The
parameter $\beta_s$ is expected to be small in the Standard Model and
can be sensitive to New Physics.  The same $b\to c \bar c s$ process
can also produce the decay $B_s^0\to J/\psi f_0(980)$, which is
another promising channel for accessing the mixing parameters, with
the clear advantage that no angular analysis is required because of
the $J^P=0^+$ quantum numbers of the $f_0(980)$.

Leading-order light-cone QCD predicts that the branching fraction
$\mathcal{B}(B_s^0\to J/\psi f_0(980)) = (3.1\pm 2.4)\times
10^{-4}$~\cite{Colangelo:2010bg}.  The ratio $\mathcal{R}_{f_0/\phi} =
\frac {\Gamma(B_s^0\to J/\psi f_0(980);f_0(980)\to \pi^+\pi^-)}
{\Gamma(B_s^0\to J/\psi \phi;\phi \to K^+K^-)}$ is expected to lie in
the range $0.2\lesssim \mathcal{R}_{f_0/\phi}\lesssim 0.5$, based on
scaling from the measurements of $D_s$ decays to $f_0$ and $\phi$
mesons~\cite{Stone:2008ak}.  Using the world-average value
$\mathcal{B}(B_s^0\to J/\psi \phi) = (1.3\pm 0.4\pm 0.2)\times
10^{-3}$~\cite{Nakamura:2010zzi}, we obtain $ 1.3\times 10^{-4}
\lesssim \mathcal{B}(B_s^0\to J/\psi f_0(980); f_0(980)\to
\pi^+\pi^-)\lesssim 3.2\times 10^{-4} $.  A recent
study~\cite{Leitner:2010fq} also shows that $\mathcal{R}_{f_0/\phi}$
can be used to estimate the S-wave pollution in the $B_s^0\to
J/\psi\phi$ analysis of $\beta_s$.

We study $B_s^0\to J/\psi f_0(980)$ in fully reconstructed $B_s^0\to
J/\psi\pi^+\pi^-$ final states using a $ 121.4\;\mathrm{fb}^{-1}$ data
sample collected at the $\Upsilon(5S)$ resonance with the Belle
detector at the KEKB collider~\cite{KEKB}.  $B_s^0$ mesons can be
produced in three $\Upsilon(5S)$ decays: $\Upsilon(5S) \to B_s^*
\bar{B}_s^*$, $B_s^*\bar{B}_s^0$ and $B_s^0 \bar{B}_s^0$ where the
$B_s^*$ mesons decay to $B_s^0 \gamma$.  The number of $B_s^{(*)}\bar
B_s^{(*)}$ pairs in the sample is measured to be $N_{B_s^{(*)}\bar
B_s^{(*)}} = (7.1 \pm 1.3) \times 10^6$ using methods described
in~\cite{cleo-fs,belle-fs}.  Production fractions are measured with
fully reconstructed $B_s^0 \to D_s^- \pi^+$ decays as described
in~\cite{:2008sc}.  We determine the fraction of $B_s^*\bar B_s^*$
pairs among all $B_s^{(*)}\bar B_s^{(*)}$ events to be
$f_{B_s^*\bar{B}_s^*} = (87.0 \pm 1.7)\%$ in our full data sample.
The number of $B_s^0$ mesons in the dominant $B_s^*\bar B_s^*$
production mode is thus $N_{B_s^0} = 2N_{B_s^{(*)}\bar B_s^{(*)}}
f_{B_s^*\bar{B}_s^*} = (1.24\pm 0.23)\times 10^7$.

The Belle detector~\cite{:2000cg} is a large-solid-angle magnetic
spectrometer that consists of a silicon vertex detector, a 50-layer
central drift chamber (CDC), an array of aerogel threshold Cherenkov
counters (ACC), a barrel-like arrangement of time-of-flight
scintillation counters (TOF), and an electromagnetic calorimeter (ECL)
comprised of CsI(Tl) crystals located inside a superconducting
solenoid coil that provides a 1.5 T magnetic field.  An iron
flux-return located outside the coil is instrumented to detect $K_L^0$
mesons and identify muons (KLM).

Charged tracks are required to originate within $0.5$ cm in the radial
direction and within 5 cm along the beam direction, with respect to
the interaction point.  Electron candidates are identified by
combining information from the ECL, the CDC $(dE/dx)$, and the ACC.
Muon candidates are identified through the track penetration depth and
hit patterns in the KLM system.  For both electrons and muons, the
identification efficiency is nearly 100\%.  Identification of charged
pions is based on the information from the CDC $(dE/dx)$, the TOF and
the ACC.  For a pion from $B_s^0\to J/\psi f_0(980)$, the
momentum-averaged identification efficiency is about 96\% with a 22\%
kaon misidentification probability.

Two oppositely charged leptons $l^+l^-$ ($l=e\;\mathrm{or}\;\mu$) and
any bremsstrahlung photons lying within 50 mrad of the $e^+$ or $e^-$
tracks are combined to form a $J/\psi$ candidate.  The invariant mass
is required to lie in the ranges
$-0.150\;\mathrm{GeV}/c^2<M_{ee(\gamma)}-
m_{J/\psi}<0.036\;\mathrm{GeV}/c^2$ or
$-0.060\;\mathrm{GeV}/c^2<M_{\mu\mu}-
m_{J/\psi}<0.036\;\mathrm{GeV}/c^2$, where $m_{J/\psi}$ denotes the
$J/\psi$ mass, and $M_{ee(\gamma)}$ and $M_{\mu\mu}$ are the
reconstructed invariant masses for $e^+e^-(\gamma)$ and $\mu^+\mu^-$,
respectively.  We combine the $J/\psi$ candidate and a $\pi^+\pi^-$
pair to form a $B_s^0$ meson.  The $\pi^+\pi^-$ and $J/\psi$ vertex
positions are required to be consistent.

Two kinematic variables are computed in the $e^+e^-$ collision rest
frame: the energy difference $\Delta E= E_B^*-E_\mathrm{beam}$, and
the beam-energy constrained mass
$M_\mathrm{bc}=\sqrt{(E_\mathrm{beam})^2- (p_B^*)^2}$, where $E_B^*$
and $p_B^*$ are the energy and momentum of the reconstructed $B_s^0$
candidates and $E_\mathrm{beam}$ is the beam energy.  To improve the
$\Delta E$ and $M_\mathrm{bc}$ resolutions, mass-constrained kinematic
fits are applied to $J/\psi$ candidates.  As $B_s^*\bar B_s^*$ pairs
are dominant in all $B_s^{(*)}\bar B_s^{(*)}$ events, we focus on the
analysis of the $\Upsilon(5S)\to B_s^*\bar B_s^*$ channel.  After an
initial loose selection, 66\% of events have multiple candidates.
From these we choose the candidate with an $M_\mathrm{bc}$ value
closest to the nominal $B_s^{*}$ mass. This requirement has an
efficiency of 90\% for the correctly reconstructed signal, according
to Monte Carlo (MC) simulations.  We then select events that lie
inside a $3\sigma$ $M_\mathrm{bc}$ signal region for $B_s^{*}\bar
B_s^{*}$ with the criterion $5.4041\;\mathrm{GeV}/c^2< M_\mathrm{bc}
<5.4275\;\mathrm{GeV}/c^2$, which rejects $\Upsilon(5S)\to
B_s^{(*)}\bar B_s$ events.  We use $\Delta E$ and the $\pi^+\pi^-$
invariant mass $M_{\pi\pi}$ to extract the signal.

To suppress two-jet-like continuum background arising from $e^+e^-\to
q\bar q\;(q=u,d,s,c)$, we require the ratio of the second to zeroth
Fox-Wolfram moments~\cite{Fox:1978vu} to be less than $0.4$.  This
requirement is optimized by maximizing the figure-of-merit
$N_\mathrm{S}/\sqrt{N_\mathrm{S}+N_\mathrm{B}}$, where $N_\mathrm{S}$
is the expected number of signal events and $N_B$ is the expected
number of background events in the ($\Delta E, M_{\pi\pi}$) signal
box.  Other major background sources are from $B\bar B$ ($B \equiv
B_s^0,\;B_d^0,\;B_u^\pm$) events with one $B$ meson decaying to a
final state with a $J/\psi$ (denoted $J/\psi X$).  We use a sample of
simulated $\Upsilon(5S)$ decays, with the most recent $B$ meson pair
production rates~\cite{Drutskoy:2010an} and all known $B\to J/\psi X$
processes, to estimate this background.  The fit region is chosen to
be $-0.1\;\mathrm{GeV} < \Delta E< 0.2 $ GeV and $M_{\pi\pi}<2.0
\;\mathrm{GeV}/c^2$.  Background from $B_d^0\to J/\psi \pi^+\pi^-$
peaks near $\Delta E = -0.14$ GeV and is outside the fit region.

A study of the $J/\psi X$ MC simulation is used to categorize these
background components according to their origins and shapes.  The
expected yields in the fit region from each source are: (a) $B_s^0\to
J/\psi \eta'$, 2.6 events, with $\eta'\to \rho^0\gamma$ in which the
photon is lost; (b) $B^+\to J/\psi (K^+,\pi^+)$, 45.3 events, which
enter the fit region after combining with a random pion; and (c) other
$J/\psi X$ sources, 240.4 events, that do not peak in $\Delta E$ and
$M_{\pi\pi}$.  There are negligible correlations between $\Delta E$
and $M_{\pi\pi}$ for (b) and (c), which are parameterized by the
product of a smooth $\Delta E$ function with a threshold $M_{\pi\pi}$
function for the two-dimensional probability density function
(PDF). The $B^+\to J/\psi K^+$ and $B^+\to J/\psi \pi^+$ background
shapes are treated separately.  For (a), the shape and yield are
obtained from a dedicated MC simulation and the measured branching
fraction~\cite{:2009usa}, where a MC-generated two-dimensional PDF is
used, since there are correlations between $\Delta E$ and $M_{\pi\pi}$
that are difficult to parameterize analytically.  The non-$J/\psi$
background is studied with data from a $J/\psi$ mass
($M_{ll(\gamma)}$) sideband defined as $2.5\;\mathrm{GeV}/c^2
<M_{ll(\gamma)}<3.4\;\mathrm{GeV}/c^2$, with the regions
$-0.200\;(-0.080)\;\mathrm{GeV}/c^2<M_{ll(\gamma)}- m_{J/\psi}
<0.048\;\mathrm{GeV}/c^2$ for $J/\psi\to e^+e^-\;(\mu^+\mu^-)$
excluded.  The shape and yield of the non-$J/\psi$ background is
obtained by fitting and counting the $J/\psi$ sideband data.  In this
procedure, the lepton identification requirements are relaxed in the
fitting to enhance the statistics.  A scale factor is used in the
counting; this factor is the MC-determined ratio of the non-$J/\psi$
yield in $J/\psi$ selection window to the yield in the $J/\psi$
sideband region.
%The real data just below $\Upsilon(4S)$ resonance are also studied and no real
%$J/\psi$'s in non-$J/\psi$ background is found.

\begin{figure}\centering
\setlength{\unitlength}{0.01\columnwidth}
\vbox{
\includegraphics[width=\columnwidth]{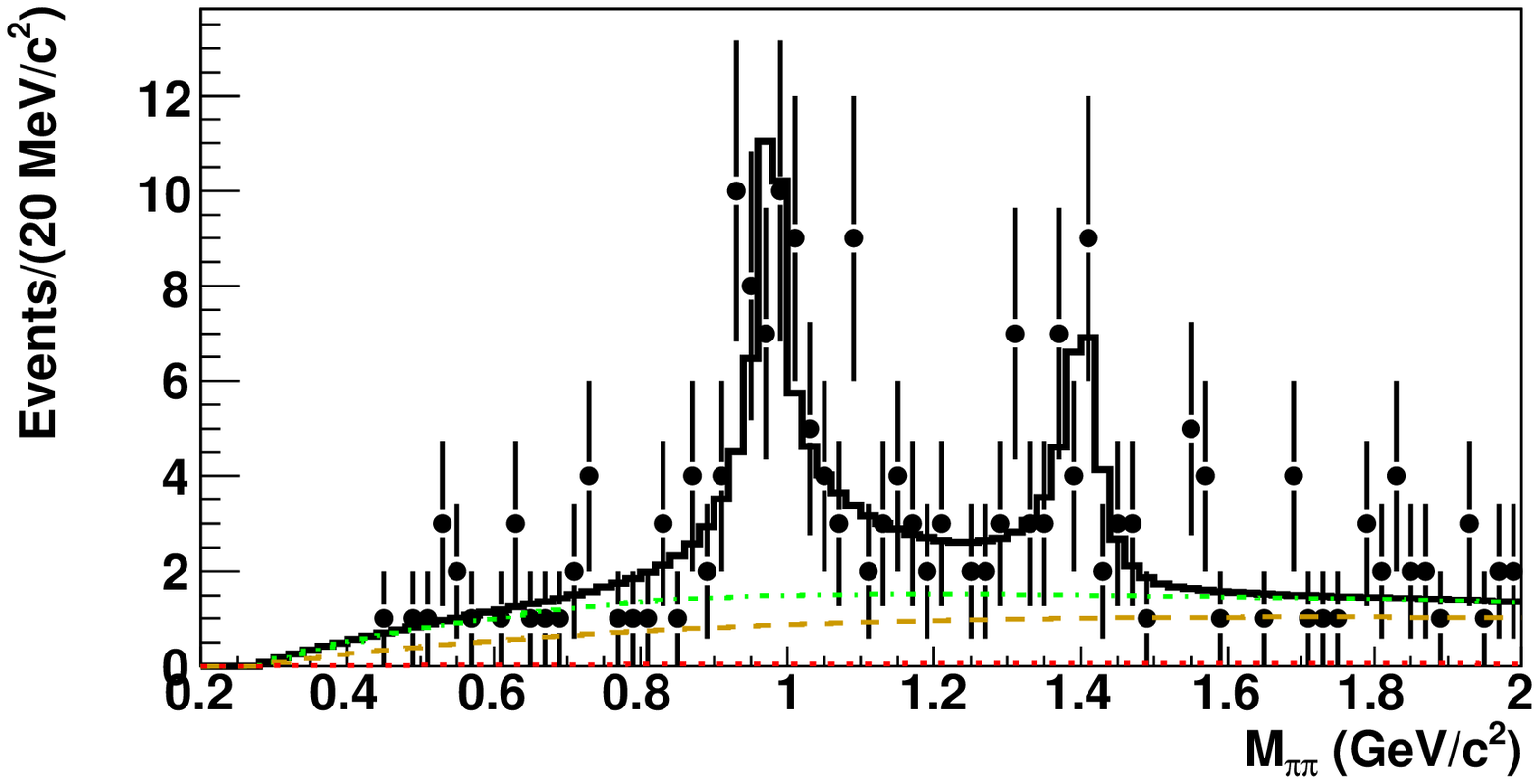}%
\begin{picture}(0,0)(0,0)
\put(-20,38){(a)}
\end{picture}
\includegraphics[width=\columnwidth]{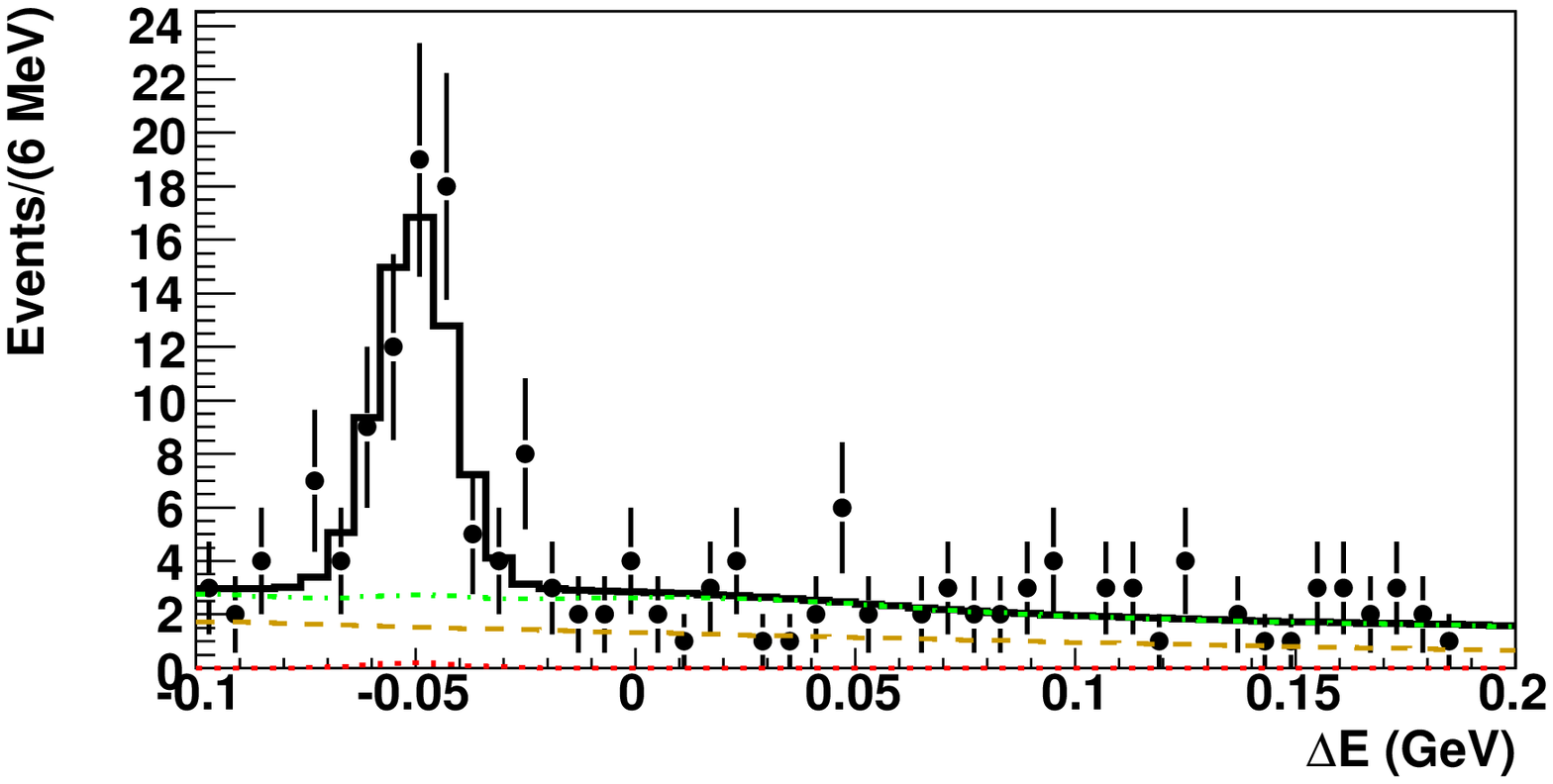}%
\begin{picture}(0,0)(0,0)
\put(-20,38){(b)}
\end{picture}
\includegraphics[width=\columnwidth]{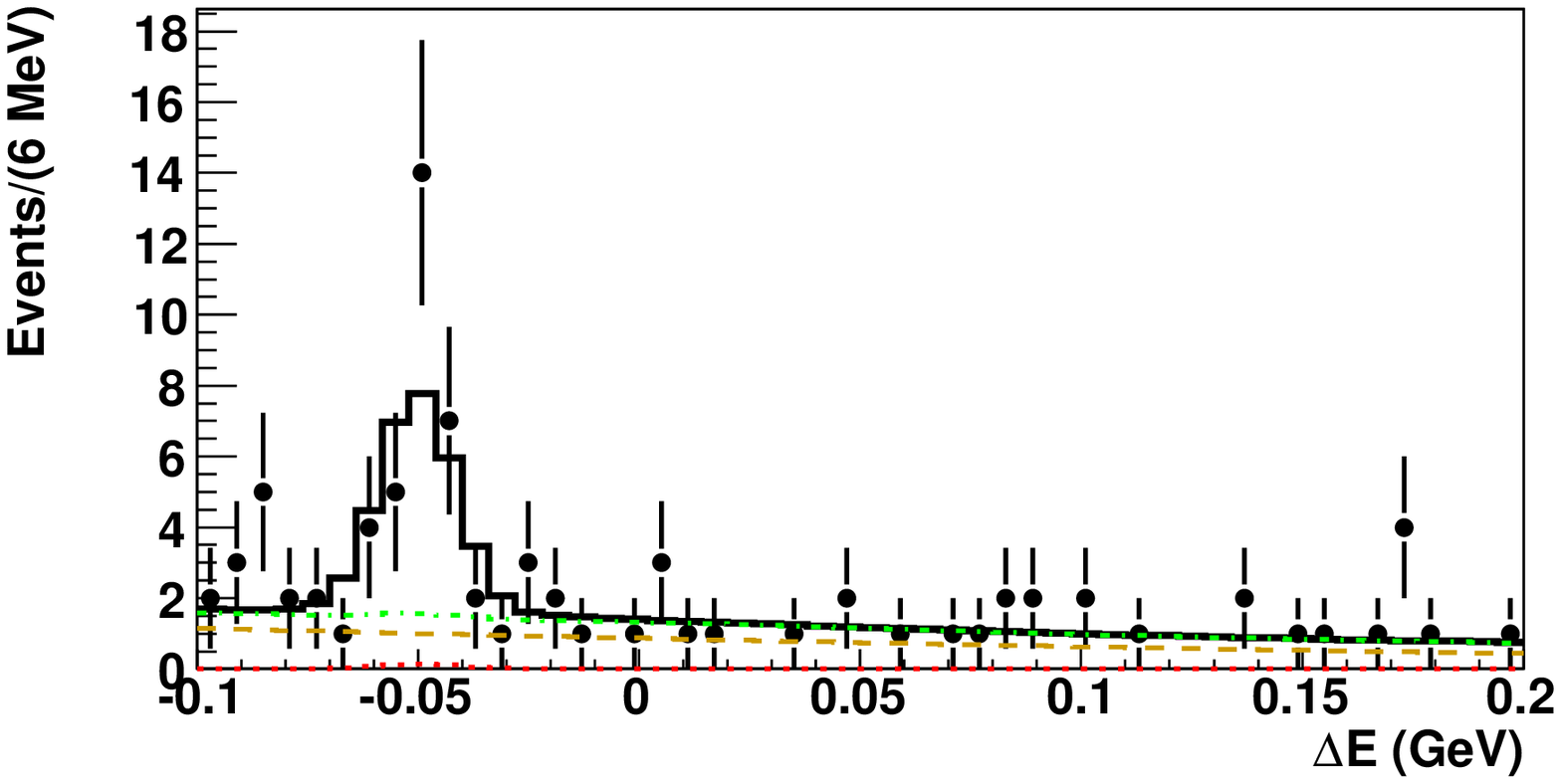}%
\begin{picture}(0,0)(0,0)
\put(-20,38){(c)}
\end{picture}
}
\caption{\label{fig:5S_scan-hde_mpipi} Data fit projections to
$M_{\pi\pi}$ for (a) $ -79.7\;\mathrm{MeV}<\Delta E <
-19.7\;\mathrm{MeV}$, and to $\Delta E$ for (b) $
0.8\;\mathrm{GeV}/c^2< M_{\pi\pi}< 1.16\;\mathrm{GeV}/c^2$ and (c) $
1.3\;\mathrm{GeV}/c^2< M_{\pi\pi} <1.5\;\mathrm{GeV}/c^2$.  The total
PDF is shown with a solid line.  The dash-dotted curves represent the
total background, the dashed curves show other $J/\psi$ background,
and the dotted curves are the non-resonant component.  }
\end{figure}

Figure~\ref{fig:5S_scan-hde_mpipi} shows the data together with the
fitting functions described below.  We find two peaks in the
$M_{\pi\pi}$ spectrum of the events in the $\Delta E$ signal region:
one for $f_0(980)$ and another around 1.4 GeV$/c^2$.  We model the
signal $M_{\pi\pi}$ PDF as a coherent sum of a Flatt\'e
function~\cite{Flatte:1976xu} for the $f_0(980)$ resonance and a
relativistic Breit-Wigner function for a second $f_X$ resonance with
mass $m_0(f_X)$ and width $\Gamma_0(f_X)$:
\begin{align}
P(M_{\pi\pi})
  = p_{J/\psi}\left| \frac{p_{J/\psi}\sqrt{M_{\pi\pi}\Gamma_1}}
  {m_0(f_0)^2-M_{\pi\pi}^2 -i(g_1\rho_1+g_2\rho_2) }\right. 
\nonumber \\
 \left. + ae^{i\theta}\frac{p_{J/\psi}\sqrt{M_{\pi\pi}\Gamma(f_X)}}
  {m_0(f_X)^2-M_{\pi\pi}^2 -im_0(f_X)\Gamma(f_X)} \right|^2 ,
\label{eqn:mpipi2r-pdf}
\end{align}
where the phase-space factors are $\rho_1=2q/M_{\pi\pi}$,
$\rho_2=2q_K/M_{\pi\pi}$, and the mass-dependent widths are $\Gamma_1
= g_1\rho_1/m_0$ and $\Gamma(f_X) =
\Gamma_0(f_X)(q/q_0)(m_0(f_X)/M_{\pi\pi})$. Here $q$ (or $q_0$) is the
pion momentum in the di-pion rest frame where the di-pion mass is
$M_{\pi\pi}$ ($m_0$), while $q_K$ is the momentum a kaon would have if
the resonance decayed to a kaon pair.  The $J/\psi$ momentum in the
$B_s^0$ rest frame, $p_{J/\psi}$, is a phase-space factor outside the
modulus, and a spin factor inside the modulus for $L=1$.  The Flatt\'e
function follows the BES parameterization~\cite{Ablikim:2004wn}, with
the parameters $m_0(f_0) = 965\pm10$ MeV$/c^2$, $g_1 = 0.165\pm
0.018\,\mathrm{GeV}^2/c^4$, and $g_2/g_1 = 4.21\pm 0.33$.  The $\Delta
E$ PDF for signal is parameterized as a sum of two Gaussians with
width calibrated using a control sample of $\Upsilon(5S)\to
B_d^{*0}\bar B_d^{*0}, B_d^0\to J/\psi K^{*0}(K^+\pi^-)$ in data.
%All the selection criteria for the control sample are the same
%except for the $K^{*0}$ selection
%and the $B_d^{*0}\overline B_d^{*0}$'s  $M_\mathrm{bc}$ signal region selection.
%A fudge factor $f_\sigma = 1.215\pm 0.063$
%is found for the $\Delta E$ width parameter,
%which will be used in the final data fit.

Contributions from the self-cross-feed (SCF) events in which one or
two pion tracks from the signal are mis-reconstructed as well as
non-resonant $B_s^0\to J/\psi\pi^+\pi^-$ events are also considered.
The SCF component is fixed to the MC value of 6.0\% of the total
signal yield in the fit region; the PDF shape is modeled with a
non-parametric histogram.  For the non-resonant background, the
$M_{\pi\pi}$ shape is obtained from a phase-space model and the
$\Delta E$ shape is the same as that of the $J/\psi f_0$ signal.

An unbinned extended maximum likelihood fit to the data is performed
using the sum of all component PDFs.  The parameters allowed to vary
in the fit are the total resonant signal yield, which includes the SCF
contribution, parameters $a$, $ \theta$, $m_0(f_X)$, $\Gamma_0(f_X)$,
the yield of the non-resonant background, and the yield of other
$J/\psi X$ background.  The yield of the $J/\psi (K, \pi)$ background
is fixed to the MC expectation.

\begin{figure}\centering
\includegraphics[width=0.5\columnwidth]{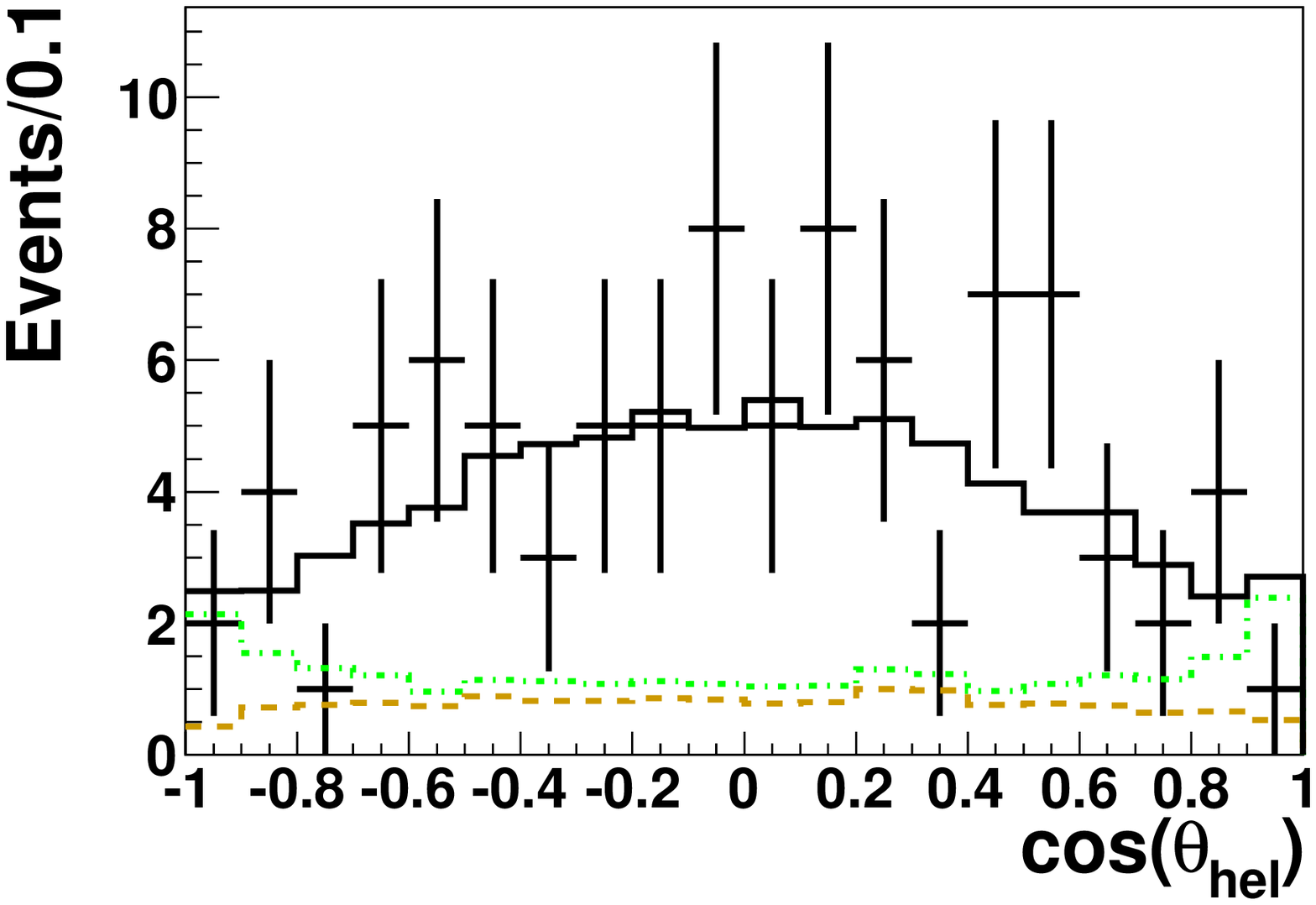}%
\includegraphics[width=0.5\columnwidth]{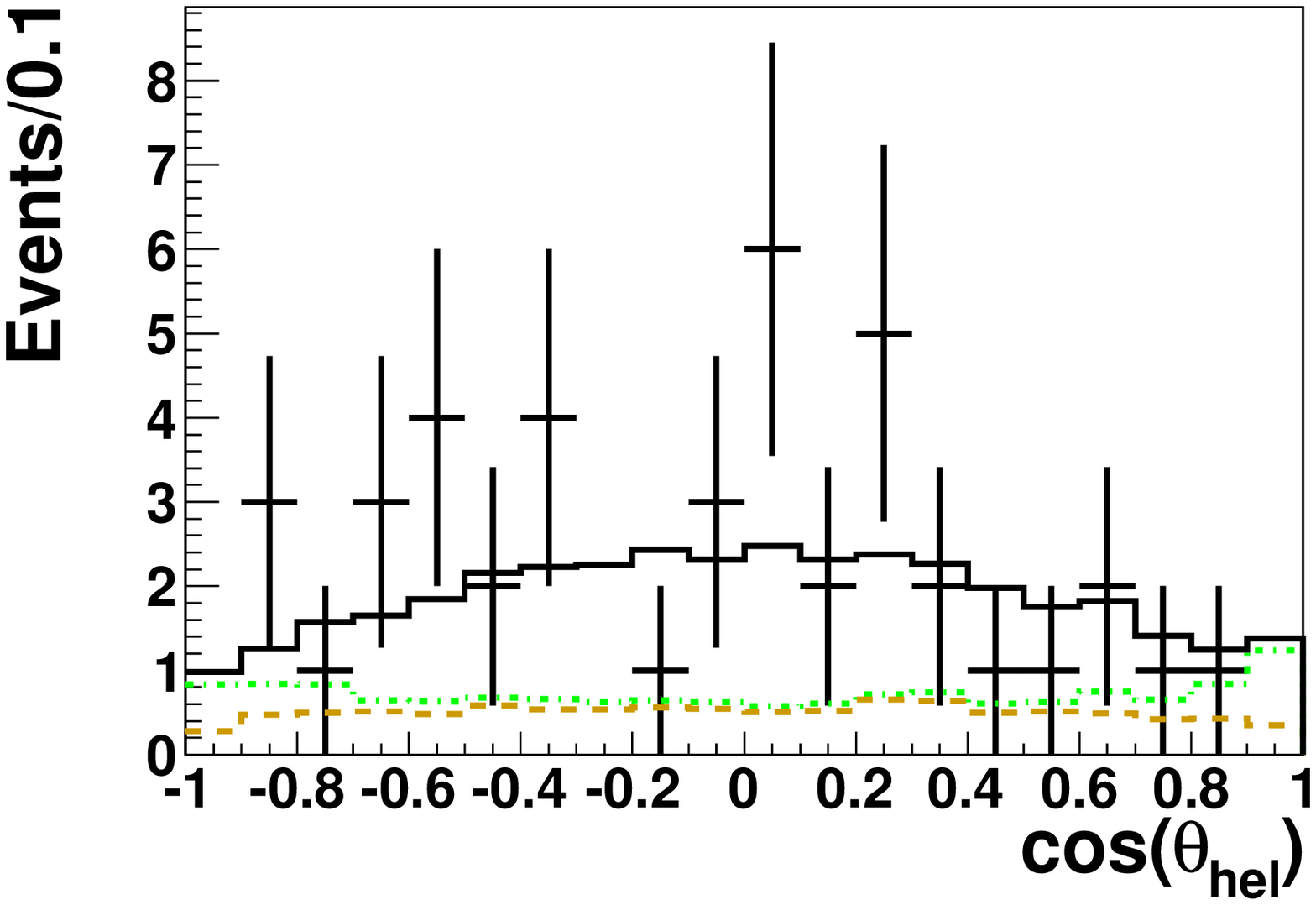}%
\caption{\label{fig:5S_scan-hhj_smd} The cosine of the angle
$\theta_\mathrm{hel}$ between $l^+$ and the direction opposite to that
of the $B_s^0$ in the $J/\psi$ rest frame.
$\cos(\theta_\mathrm{hel})$ is projected in the $\Delta E$ signal
region and $f_0(980)$ and $f_0(1370)$ signal regions as $
0.8\;\mathrm{GeV}/c^2< M_{\pi\pi}< 1.16\;\mathrm{GeV}/c^2$ (left) and
$ 1.3\;\mathrm{GeV}/c^2< M_{\pi\pi}< 1.5\;\mathrm{GeV}/c^2$ (right).
The expected distributions from the fit assuming a longitudinally
polarized $J/\psi$, which would result from a $B_s^0\to J/\psi +
\mathrm{scalar}$ decay, are superimposed.  The curves follow the
convention in Fig.~\ref{fig:5S_scan-hde_mpipi}.  }
\end{figure}

We obtain $98\pm 15$ resonant events corresponding to
Eq.(\ref{eqn:mpipi2r-pdf}), where $63^{+16}_{-10}$ are from $B_s^0\to
J/\psi f_0(980)$ and $19^{+6}_{-8}$ are from $B_s^0\to J/\psi f_X$,
and the rest are from the interference.  The yield for each signal
component is calculated using the amplitude squared of that component
divided by the coherent sum of all amplitudes, with statistical errors
obtained from error propagation using the covariant matrix of relevant
parameters.  The amplitude and phase parameters are determined to be
$a=0.47\pm0.10$ and $\theta=1.63\pm0.98$ rad.  The fitted mass and
width for the $f_X$ component are
$m_0=1.405\pm0.015^{+0.001}_{-0.007}$ GeV$/c^2$ and
$\Gamma_0=0.054\pm0.033^{+0.014}_{-0.003}$ GeV, where the first error
is statistical and the second is systematic.  These results are
consistent with the $f_0(1370)$ parameters listed in the
PDG~\cite{Nakamura:2010zzi}.  Henceforth, we refer to this
contribution as $B_s^0 \to J/\psi f_0(1370)$; however, the possibility
of other scalar resonance contributions in this region cannot be
excluded with the present statistics.
%The $f_0(1370)$ parameters are in marginal agreement with
% the $f_0(1370)$ parameters
%($m=1.449\pm 0.013\;\mathrm{GeV}/c^2,$ $\Gamma=0.126\pm 0.025\;\mathrm{GeV}/c^2$)
%obtained in Belle's $B^+\to K^+\pi^+\pi^-$ analysis~\cite{Garmash:2005rv}.
The obtained non-resonant yield is $4 \pm 12$, consistent with zero.
The yield of other $J/\psi$ background $262\pm 23$ is consistent with
the MC estimate of 240 events.  The $J/\psi$ helicity distributions,
shown in Fig.~\ref{fig:5S_scan-hhj_smd}, are consistent with a
longitudinally polarized $J/\psi$ in both $f_0(980)$ and $f_0(1370)$
signal regions, as expected for scalar $\pi\pi$ resonances.

The signal yields, branching fractions, and significances including
systematic uncertainties are listed in Table~\ref{tab:5S_scan-bf0}.
The significance is calculated from the log likelihood difference
for two parameters in the $f_0(980)$ case and four parameters in the
$f_0(1370)$ case, when the corresponding signal amplitude is set to
zero.

%The total number of 
%$B_s^{*}\bar B_s^{*}$ pairs produced is $6.44 \times 10^{-6}$, obtained by using
%the luminosity $L$, cross section $\sigma_{b\bar b}^{\Upsilon(5S)} = (0.302\pm
%0.014)\;\mathrm{nb}$, fraction of $B_s^{(*)}\bar B_s^{(*)}$ pairs 
%$f_s =(19.5^{+3.0}_{-2.3})\%$~\cite{Nakamura:2010zzi}, and $f_{B_s^*\bar B_s^*}$ at the
% $\Upsilon(5S)$ energy.

%\begin{widetext}
\begin{table*}
\caption{\label{tab:5S_scan-bf0}
Summary of signal yields, significances,
and product branching fractions $\mathcal{B}(B_s^0\to J/\psi F; F\to\pi^+\pi^-)$,
where $F=f_0(980)\ \mathrm{or}\ f_0(1370)$. }
\begin{ruledtabular}
\begin{tabular}{lccc}
Mode & Yield  & Significance & $\mathcal{B}(B_s^0\to J/\psi F; F\to\pi^+\pi^-)$ \\ \hline
$B_s^0\to J/\psi f_0(980)$  & $63^{+16}_{-10}$ &  8.4$\sigma$ & \bsjf0BF \\
$B_s^0\to J/\psi f_0(1370)$ & $19^{+6}_{-8}$   &  4.2$\sigma$ & \bsjfxBF \\
\end{tabular}
\end{ruledtabular}
\end{table*}
%\end{widetext}

\begin{table}
\caption{\label{tab:errBF_syst}
Relative systematic errors (in \%) for 
$\mathcal{B}(B_s^0\to J/\psi F; F\to\pi^+\pi^-)$.}
\begin{ruledtabular}
\begin{tabular}{lcc}
Source  &  $\mathcal{B}(F= f_0(980))$ & $\mathcal{B}(F=f_0(1370))$\\ \hline
$\Delta E$ shape   & 1.0              & $+0.8,-0.6$ \\
$f_0(980)$ shape   &  $+12.4, - 13.6$ &  $+8.9, -5.3$ \\
Background parameters & $+1.9,-1.7$   & $+1.0,-0.4$ \\  \hline
Track reconstruction  & \multicolumn{2}{c}{1.3}  \\
Lepton identification & \multicolumn{2}{c}{2.6} \\
Pion identification   & \multicolumn{2}{c}{1.6}  \\
$\mathcal{B}(J/\psi \to ll)$ & \multicolumn{2}{c}{0.7} \\
$f_{B_s^*\bar B_s^*}$ &  \multicolumn{2}{c}{2.0} \\
$N_{B_s^{(*)}\bar B_s^{(*)}} $ & \multicolumn{2}{c}{$+22.4, -15.5$} \\  \hline
Total & $+26.0, -21.1 $ & $+24.5,-16.8$ \\
\end{tabular}
\end{ruledtabular}
\end{table}

Contributions to the systematic error are obtained by varying each
fixed parameter by its error and are summarized in
Table~\ref{tab:errBF_syst}.  Apart from the $N_{B_s^*\bar B_s^*}$
normalization, the largest systematic effect arises from the
uncertainties of the Flatt\'e parameters for the $f_0(980)$ lineshape,
where the parameters are varied according to errors in
\cite{Ablikim:2004wn}.  For the signal $\Delta E$ shape, the error on
the mean value is determined from the beam energy calibrated with
$\Upsilon(5S)\to \Upsilon(1S)\pi\pi$ and $B_s^0\to D_s\pi$, and the
error on the width is determined from the control sample.  The yields
of the $B_s^0 \to J/\psi \eta'$ and $B^+\to J/\psi (K^+,\pi^+)$
components are varied according to the experimental errors on their
branching fractions.  Finally, the non-$J/\psi$ background parameters
are varied according to the results of the $J/\psi$ sideband study.
Other background shape uncertainties are negligible.

In summary, we report the first observation of $B_s^0\to J/\psi
f_0(980)$ and the first evidence for $B_s^0\to J/\psi f_0(1370)$.  The
measured $B_s^0\to J/\psi f_0(980)$ branching fraction is in agreement
with the estimate $ 1.3\times 10^{-4} \lesssim \mathcal{B}(B_s^0\to
J/\psi f_0(980); f_0(980)\to \pi^+\pi^-)\lesssim 3.2\times 10^{-4} $.
The signal for $B_s^0\to J/\psi f_0(1370)$ has a
significance of $4.2 \sigma$.  This mode represents a new $CP$ channel
that can be used to study $B_s^0$ mixing properties.

Note added: While preparing the final version of this manuscript, we
became aware of arXiv:1102.0206, which reports similar results for
$B_s^0\to J/\psi f_0(980)$.

We thank the KEKB group for excellent operation of the
accelerator, the KEK cryogenics group for efficient solenoid
operations, and the KEK computer group and
the NII for valuable computing and SINET3 network support.  
We acknowledge support from MEXT, JSPS and Nagoya's TLPRC (Japan);
ARC and DIISR (Australia); NSFC (China); MSMT (Czechia);
DST (India); MEST, NRF, NSDC of KISTI, and WCU (Korea); MNiSW (Poland); 
MES and RFAAE (Russia); ARRS (Slovenia); SNSF (Switzerland); 
NSC and MOE (Taiwan); and DOE (USA).

\end{document}